\documentstyle[aps,pre,epsfig]{revtex}
\begin{document}
\draft 
\title{Freak Waves in Random Oceanic Sea States }
\author{ Miguel Onorato, Alfred R. Osborne, Marina Serio and Serena
Bertone}
\address{Dip. di Fisica Generale, Universit\'{a} di Torino, Via
Pietro Giuria 1, 10125 Torino, Italy}
\date{\today}
\maketitle
\begin{abstract}
Freak waves are very large, rare events in a random ocean wave train. 
Here we study the  numerical generation of freak waves in a
random sea state characterized by the JONSWAP power spectrum. We
assume, to cubic order in nonlinearity, that the wave dynamics are
governed by the nonlinear Schroedinger (NLS) equation. We identify
two parameters in the power spectrum that control the
nonlinear dynamics: the Phillips parameter
$\alpha$ and the enhancement coefficient $\gamma$. We discuss how freak waves in
a random sea state are more likely to occur for large values of
$\alpha$ and $\gamma$. Our results are supported by extensive
numerical simulations of the NLS equation with random initial
conditions. Comparison with linear simulations are also reported.
\end{abstract}
\vspace{0.5cm}
%
Freak waves are extraordinarily large water waves whose heights exceed by a
factor of 2.2 the significant wave height of a measured wave train \cite
{FREA}. The mechanism of freak wave generation has become an issue of
principal interest due to their potentially devastating effects on offshore
structures and ships. In addition to the formation of such waves in the
presence of strong currents \cite{WHI} or as a result of a simple chance
superposition of Fourier modes with coherent phases, it has recently been
established that the nonlinear Schroedinger (NLS) equation can describe many
of the features of the dynamics of freak waves which are found to arise as a
result of the nonlinear self-focusing phenomena \cite{TDF,DT,PER}. The
self-focusing effect arises from the Benjamin-Feir instability \cite{BEF}: a
monochromatic wave of amplitude, $a_0$, and wave number, $k_0$,
modulationally perturbed on a wavelength $L=2 \pi/ \Delta k$, is unstable
whenever $\Delta k/(k_0\varepsilon)<2\sqrt{2}$, where $\varepsilon$ is the
steepness of the carrier wave defined as $\varepsilon$=$k_0a_0$. The
instability causes a local exponential growth in the amplitude of the wave
train. This result is established from a linear stability analysis of the
NLS equation \cite{YUEN} and has been confirmed, for small values of the
steepness, by numerical simulations of the fully nonlinear water wave
equations \cite{PER,TUL} (for high values of steepness wave breaking, which
is clearly not included in the NLS model, can occur).  Moreover, it is known
that small-amplitude instabilities are but a particular case of the much
more complicated and general analytical solutions of the NLS equation
obtained by exploiting its integrability properties via Inverse Scattering
theory in the $\theta$-function representation \cite{KOT,TRACY}.

Even though the above results are well understood and robust from a physical
and mathematical point of view, it is still unclear how freak waves are
generated via the Benjamin-Feir instability in more realistic oceanic
conditions, i.e. in those characterized not by a simple monochromatic wave
perturbed by two small side-bands, but instead by a complex spectrum whose
perturbation of the carrier wave cannot be viewed as being small.
Furthermore, the focus herein is not to attempt to model ocean waves but
instead to study leading order effects using the nonlinear Schroedinger
equation, as suggested by \cite{TDF,DT,PER}. Research at higher order
suggests that the results given herein are indicative of many physical
phenomena in the primitive equations \cite{PER,TUL}.

\begin{figure}[htb]
\centerline{\epsfig{figure=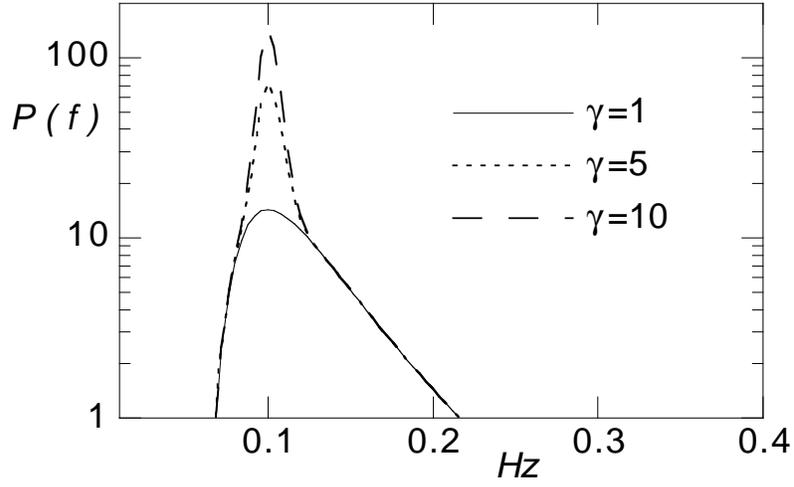,width=0.6\textwidth,angle=0}}
\caption{The JONSWAP spectrum for $\gamma$=1 (dashed line), $\gamma$=5
(dotted line), $\gamma$=10 (solid line) with $f_0$=0.1 Hz and $\alpha$%
=0.0081.}
\label{fig pm_jon}
\end{figure}

In this Letter our attention is focused on freak wave generation in
numerical simulations of the NLS equation where we assume initial conditions
typical of oceanic sea states described by the JONSWAP power spectrum (see,
e.g. \cite{KOM}): 
\begin{equation}
{P(f)= \frac {\alpha} {f^5} \exp \bigg[-\frac{5} {4} \bigg(\frac{f_0} {f}%
\bigg)^4\bigg]\gamma^ {\exp\bigg[-\frac{(f-f_0)^2} {2\sigma_0^2f_0^2}\bigg]} 
}  \label{jonswap}
\end{equation}
where $\sigma_0$=0.07 if $f\le f_0$ and $\sigma_0$=0.09 if $f> f_0$. Our use
of the JONSWAP formula is based upon the established result that developing
storm dynamics are governed by this spectrum for a range of the parameters 
\cite{{KOM}}. The constants $\alpha$, $\gamma$ and $\sigma_0$ were
originally obtained by fitting experimental data from the international
JONSWAP experiment conducted during 1968-69 in the North Sea. Here $f_0$ is
the dominant frequency, $\gamma$ is the ``enhancement'' coefficient and $%
\alpha$ is the Phillips parameter. For $\gamma$=1 and $\alpha$=0.0081 the
spectrum reduces identically to that of Pierson and Moskowitz \cite{{KOM}}
which describes a fully developed sea state, i.e. one which has evolved over
infinite time and space. For $f=f_0$, the expression $E$ = $\gamma^ {\exp[...%
]}$ is effectively equal to the value of $\gamma$ and, as $f$ moves away
from $f_0$ in either direction, $E$ tends rapidly to unity. Therefore as $%
\gamma$ increases, the spectrum becomes higher and narrower around the
spectral peak. In Fig. \ref{fig pm_jon} we show the JONSWAP spectrum for
different values of $\gamma$ ($\gamma$=1, 5, 10) for $f_0$=0.1 $Hz$ and $%
\alpha$=0.0081.

The major finding we would like to discuss herein is that as $\gamma$ and $%
\alpha$ grow, the nonlinearity becomes more important and the probability of
the formation of freak waves increases. Our results have been achieved by
considering the NLS equation as the $simplest$ nonlinear evolution equation
for describing deep-water wave trains. We have performed numerical
simulations using the JONSWAP spectrum to determine the initial conditions.
Since the analytical form of the spectrum is given as a function of
frequency, the analysis is carried out by considering the so called
time-like NLS equation (TNLS) (for the use of time-like equations in water
waves see, e.g., \cite{LOM,MEI,OSBPET}) which describes the evolution of the
complex envelope $A$ in deep water waves: 
\begin{equation}
{A_x + i \bigg(\frac{\Delta \omega} {\omega_0} \bigg)^2 A_{t t}+ i
\varepsilon^2 \mid A\mid^2 A =0, }  \label{NLS_ADIM}
\end{equation}
where dimensional quantities denoted with primes have been scaled according
to: $A=a_0 A^{\prime}$, $x=x^{\prime}/k_0$ and $t=t^{\prime}/\Delta \omega$
with $1/\Delta \omega$ a characteristic time scale of the envelope which
corresponds to the width of the frequency spectrum. Eq. (\ref{NLS_ADIM})
solves a {\it boundary value problem}: given the temporal evolution $A(0,t)$
at some location $x=0$, eq. (2) determines the wave motion over all space, $%
A(x,t)$.

At this point, it is instructive to introduce a parameter that estimates the
influence of the nonlinearity in deep water waves. This parameter, which is
a kind of ``Ursell'' number \cite{OSBPET}, can be obtained as the ratio of
the nonlinear and dispersive terms in the TNLS equation: 
\begin{equation}
{Ur=\bigg(\frac{\varepsilon} {\Delta \omega/\omega_0} \bigg)^2 .}
\label{URSELL}
\end{equation}
When $Ur<<1$ waves are essentially linear and their dynamics can be
expressed as a simple superposition of sinusoidal waves. For $Ur\geq1$, the
dynamics become nonlinear and the evolution of the wave train is likely
dominated by envelope solitons or unstable mode solutions such as those
studied by Yuen and coworkers \cite{YUEN}.

\begin{figure}[htb]
\centerline{\epsfig{figure=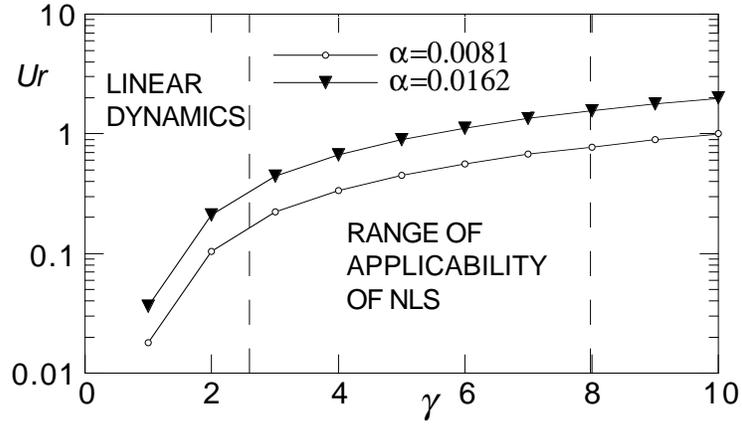,width=10cm,angle=0}}
\caption{The Ursell number as a function of $\gamma$ for the JONSWAP
spectrum.}
\label{fig ursell}
\end{figure}

Many aspects of the importance of the nonlinearity can be addressed by
computing $Ur$ from the spectrum (\ref{jonswap}). In Fig. \ref{fig ursell}
we show the Ursell number as a function of the parameter $\gamma $ for $%
\alpha $=0.0081 and $\alpha $=0.0162. In the construction of the plot an
estimation of $\varepsilon $ and $\Delta \omega /\omega _{0}$ needs to be
given. The steepness $\varepsilon $ has been estimated as the product of the
wave number, $k_{0}$, of the carrier wave with a characteristic wave
amplitude which we compute as the significant wave height (the mean of the
highest 1/3 wave heights in a wave train), $H_{s}$, divided by 2. $\Delta %
\omega $ is a measure of the width of the spectrum and it has been estimated
as the half-width at half-maximum. From the plot it is evident that for the
Pierson-Moskowitz spectrum ($\gamma =1$) the Ursell number is quite small:
this indicates that dispersion dominates nonlinearity. Formally, the NLS
equation is derived assuming that the spectrum is narrow banded and the
steepness is small. It has to be pointed out that 
for small values of $\gamma $ ($\gamma =1,2$) the
spectrum is not narrow banded; as $\gamma $ increases the spectrum becomes
narrower ($\Delta \omega /\omega _{0}$ $\simeq $0.2 or less), suggesting
that the NLS equation is more appropriate. For large values of $\gamma $ the
mean steepness increases; for $\gamma =8$, $\alpha =0.01$ the steepness is
equal to 0.16, therefore the equation is no longer valid and higher order
terms in steepness are required. In Fig. \ref{fig ursell} we have placed
vertical lines at $\gamma =2.5$ and $\gamma =8$ to indicate the region in
which the NLS equation is applicable.

When the spectral width  becomes large, one expects results which are
somewhat out of the range of applicability of the NLS equation. As pointed
out in a number of papers \cite{TD2,TKDV} the main defect in the NLS
equation concerning the narrow-band approximation, arises from the fact that
linear dispersion is not at high enough order. Reference \cite{TKDV}
proposes an equation that includes all the terms in the linear dispersion
relation. The equation (eq. (1) in their paper), which is basically the NLS
equation with the full linear dispersion relation of the primitive
equations, ``reproduces exactly the conditions for nonlinear
four-wave resonance even for bandwidth greater than
unity''. In the linear limit the equation is exact. In our numerical
simulations with the Pierson Moskowitz spectrum ($\gamma =1$, 
$\alpha=0.0081$ ) we have used
both the NLS equation and its modified form (eq. (1) in \cite{TKDV}). In
this specific case the two equations give basically the same results:
nonlinearities are weak (Ursell number=0.03, see Fig. \ref{fig ursell}) and
the dynamics are basically linear: the correction in the linear dispersion
relation does not essentially alter the value of the maximum simulated wave
amplitudes. The results of these tests have convinced us that, for the
important range $\gamma =2.5-8$, the simpler NLS eqution is a valid approach
for studying many of the properties of rogue waves.

The influence of the parameter $\alpha$ consists in increasing the energy
content of the time series and, therefore as $\alpha$ increases, the wave
amplitude and consequently the wave steepness also increase. If $\alpha$
doubles, the steepness increases by a factor of $\sqrt{2}$ and the Ursell
number by a factor of 2 since the spectral width remains constant. From this
analysis we expect that large amplitude freak waves (large with respect to
their significant height) are more likely to occur when $\gamma$ and $\alpha$
are both large.

\begin{figure}[htb]
\centerline{\epsfig{figure=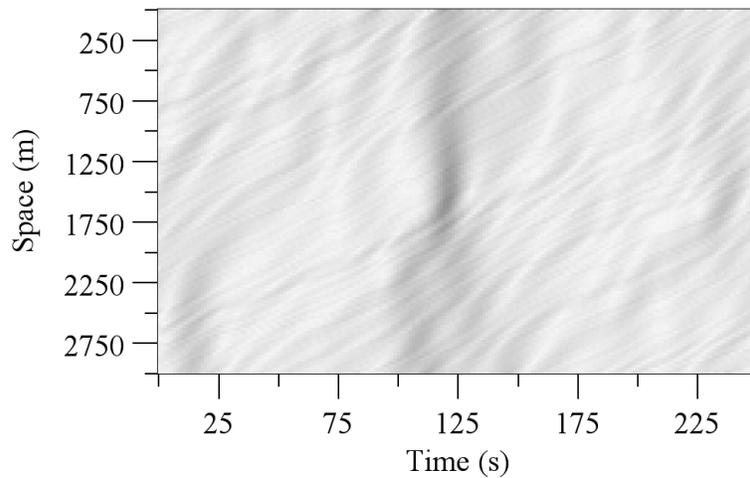,width=10cm,angle=0}}
\caption{Nonlinear Schroedinger space-time evolution of a random wave field
using the JONSWAP spectrum with $\gamma$=4, $\alpha$=0.0081. For details
refer to the text. Gray scale ranges from 0 m (white) to 12.8 m (dark).}
\label{fig fieldNLS}
\end{figure}

\begin{figure}[htb]
\centerline{\epsfig{figure=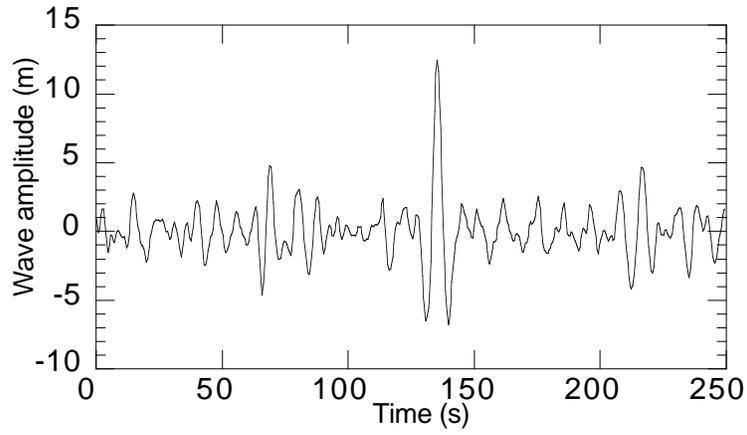,width=10cm,angle=0}}
\caption{Free surface elevation $\zeta(t)$ at $x=1550$ m obtained from Fig.
3.}
\label{fig freak}
\end{figure}

We now consider numerical simulations of eq. (\ref{NLS_ADIM}) which have
been computed using a standard split-step, pseudo-spectral Fourier method 
\cite{LOM}. Initial conditions for the free surface elevation $\zeta (0,t)$
have been constructed as the following random process \cite{OSB_EF}: 
\begin{equation}
{\zeta (0,t)=\sum_{n=1}^{N}C_{n}\cos (2\pi f_{n}t-\phi _{n}),}
\label{surface}
\end{equation}
where $\phi _{n}$ are uniformly distributed random numbers on the interval $%
(0,2\pi )$, and $C_{n}=\sqrt{2P(f_{n})\Delta f_{n}}$, where $P(f)$ is the
JONSWAP spectrum given in (\ref{jonswap}). For the computational domain
considered, it was checked that the shape of the JONSWAP spectrum has been
not altered during the evolution of the NLS equation. It was found that
during the evolution there is a continuos exchange of energy among the
frequencies around the peak. If the numerical time series at a fixed spatial
location is split into shorter time series in order to compute an average
spectrum (this procedure is usually adopted when dealing with experimental
time series), the shape of the spectrum is preserved with high accuracy, at
least for $\gamma <8$. In Fig. \ref{fig fieldNLS} we show an image of
smoothed contours of a space-time field of $|A|$ from a numerical simulation
of TNLS obtained with $\gamma $=4. The dominant frequency and the Phillips
parameter of the initial wave train were set respectively to 0.1 Hz and $%
\alpha $=0.02. A large amplitude wave appears in the simulation and in order
to better visualize it, in Fig. \ref{fig freak} we show a time series of the
free surface $\zeta (t)$ at $x=1550$ m obtained using the following
relation: 
\begin{equation}
{\zeta (t)=(A(t)e^{i2\pi f_{0}t}+c.c.)/2.}  \label{rel_surf_env}
\end{equation}
where $c.c.$ denotes complex conjugate. A freak wave of about $18.5$ m in a
random wave train with significant wave height of $H_{s}=6.9$ m is evident
at time $t=140$ s. We point out that the same simulation, not reported here
for brevity, with exactly the same initial conditions, has been performed
after deleting the nonlinear term in the TNLS equation, i.e. the term $\mid
A\mid ^{2}A$ was set identically to zero. No waves fulfilling the freak wave
threshold ($H>2.2Hs$) were found. In this linear simulation, the
Benjamin-Feir instability cannot occur and ''freak'' waves can occur only
via a simple superposition of Fourier modes. This result indicates that, in
our simulations, nonlinearity plays an important role in the dynamics of
freak wave generation.

In order to give additional quantitative results we have performed more that
300 simulations of the TNLS equation. The simulations have been performed in
dimensional units in the following way. An initial time series of 250
seconds has been computed from the JONSWAP spectrum for different values of $%
\alpha $ (from $\alpha $=0.0081 to $\alpha $=0.02) and $\gamma $ ($\gamma $%
=1, $\gamma $=4 and $\gamma $=10 ). To increase the number of statistical
events we made computer runs with 10 different sets of random phases, $\phi
_{n}$. The time series were then evolved according to the TNLS for a
distance of 10 km, saving the output every 10 m. From an experimental point
of view this approach corresponds to setting 1000 probes along the wave
propagation direction (one every 10 m) and measuring each time series for
250 s at a sampling frequency of 2.05 Hz. The significant wave height, $H_{s}
$, of each realization has been computed; the highest wave, $H_{max}$, has
been found and the ratio $H_{max}/H_{s}$ has been determined. In order to
verify that, in all the simulations performed, our findings are really a
consequence of the nonlinear dynamics, we have also computed exactly the
same simulations using the linear version of the TNLS equation. The results
are summarized in Fig. \ref{fig gamma=1}, \ref{fig gamma=4}. Fig. \ref{fig
gamma=1} corresponds to $\gamma =1$. A horizontal line at $H_{max}/H_{s}=2.2$
indicates the threshold that arbitrarily discriminates the height of rogue
waves. For the Pierson Moskowitz spectrum ($\gamma =1$ and $\alpha =0.0081$)
only one realization of the 10 considered shows a ''rogue'' wave with $%
H_{max}/H_{s}=2.25$. For higher values of $\alpha $ only a few of the
realizations show waves with $H_{max}/H_{s}$ slightly greater than 2.2. From
the plot it is clear that the effects of nonlinearities are rather small for
this case ($\gamma =1$). Among all the 50 linear simulations performed with $%
\gamma =1$, we have encoutered a number of large amplitude waves but none
exceeds 2.2 $H_{s}$.

For $\gamma$=4, see Fig. (\ref{fig gamma=4}), the physical  picture becomes
much more interesting: while in the linear simulations there are no freak
waves, 50$\%$ of the nonlinear simulations performed show at least one freak
wave.

For $\gamma$=10 the picture is qualitatively the same and therefore the
graphs are not reported. There is clear evidence that increasing $\gamma$
increases the probability of freak wave occurrences; high values of $\gamma$
do not, however, guarantee the presence of a giant wave. The local
properties of the wave trains are presumably of fundamental importance for
understanding the formation of freak waves: it may happen
that the Benjamin-Feir instability mechanism is satisfied only in a small
temporal portion of the full wave train, giving rise to a local instability
and therefore to the formation of a freak wave.

From a physical point of view, we are aware of the fact that the NLS
equation overestimates the region of instability and the maximum wave
amplitude with respect to higher order models \cite{DYS}, especially for $%
\varepsilon$ greater than 0.1. Furthermore it is well known that the NLS
equation is formally derived from the Euler equations under the assumption
of a narrow-banded process. Nevertheless, in spite of these deficiencies in
the NLS equation, we believe that our results provide new important physical
insight into the generation of freak waves. Simulations with higher order
models \cite{DYS} or directly with the fully nonlinear equations of motion
will be required in order to confirm these results. Wave tank experiments
will also be very useful in this regard.

Another issue that has to be taken into account for future work is
directional spreading: it is well known that sea states are not fully
unidirectional and directionality can play an important role in the dynamics
of ocean waves. In a recent paper \cite{OSB00} we have considered simple
initial conditions using the NLS equation in 2+1 dimensions and we have
found the ubiquitous occurrence of freak waves. Whether the additional
directionality in the JONSWAP spectrum changes our statistics is still an
open question; at the same time we are confident that our results can apply
to the case in which the spectrum is quasi-unidirectional. In particular, as
recently suggested \cite{DON}, the so called ``energetic swells'', which
correspond to the early stage of swell development (still characterized by a
highly nonlinear regime), are described by high values of $\gamma$ and $%
\alpha$. Therefore, these sea states are candidates for the occurrence of
freak waves. In such conditions 1+1 NLS represents a {\it good} evolution
equation. %
%

M. O. was supported by a Research Contract from the Universit\'{a} di
Torino. This work was supported by  the Office of Naval Research of the
United States of America (T. F. Swean, Jr. and T. Kinder) and by the Mobile
Offshore Base Program of ONR (G. Remmers). Consortium funds and Torino
University funds (60 \%) are also acknowledged.

\newpage

\begin{figure}[htb]
\centerline{\epsfig{figure=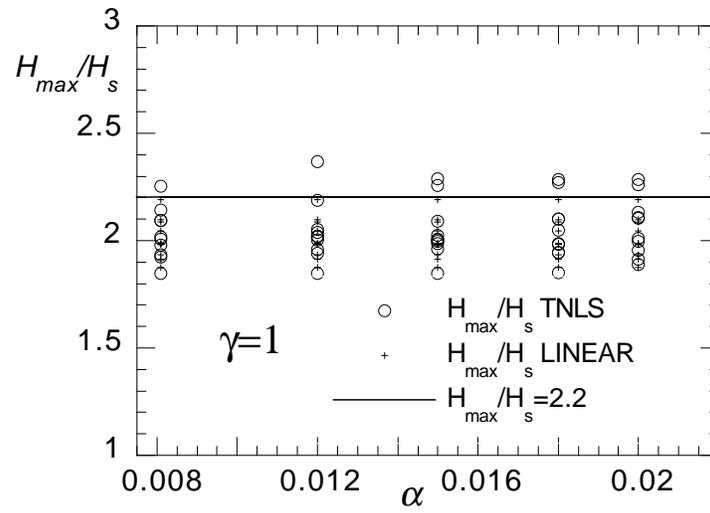,width=10cm,angle=0}}
\caption{$H_{max}/H_s$ as a function of $\alpha$ for $\gamma=1$. Circles and
crosses corresponds respectively to the nonlinear (TNLS) and linear
simulations. For each value of $\alpha$, 10 different realizations
corresponding to 10 different sets of random numbers have been performed.}
\label{fig
gamma=1}
\end{figure}
\begin{figure}[htb]
\centerline{\epsfig{figure=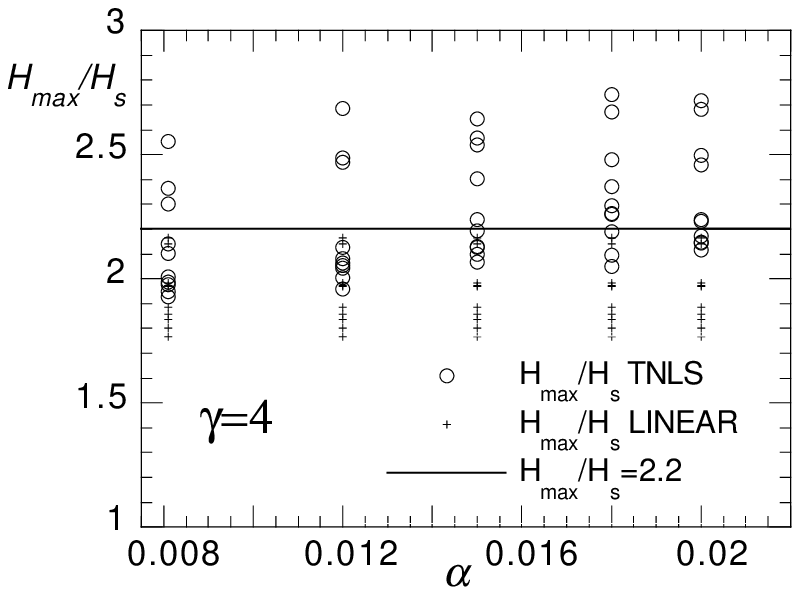,width=10cm,angle=0}}
\caption{$H_{max}/H_s$ as a function of $\alpha$ for $\gamma=4$. Circles and
crosses corresponds respectively to the nonlinear (TNLS) and linear
simulations. For each value of $\alpha$, 10 different realizations
corresponding to 10 different sets of random numbers have been performed.}
\label{fig
gamma=4}
\end{figure}
\end{document}